\documentclass[useAMS,usenatbib]{mnras}

\usepackage{amsfonts,amsmath,amssymb,mathrsfs}
\usepackage{graphicx,epsf,epsfig}
\usepackage{bm}
\usepackage{longtable}
\usepackage[usenames,dvipsnames]{xcolor}
\usepackage{multirow}

\usepackage{graphicx}
\usepackage{hyperref}
\usepackage{times}
\usepackage{color}
\usepackage{breakurl}
\usepackage{mathrsfs}

\def\be{\begin{equation}}
\def\ee{\end{equation}}
\def\bea{\begin{eqnarray}}
\def\eea{\end{eqnarray}}

\title[Static and rotating white dwarfs at finite temperatures]{Static and rotating white dwarfs at finite temperatures}

\author[K.~Boshkayev]{K.~Boshkayev,$^{1,2}$\thanks{kuantay.boshkayev@nu.edu.kz, kuantay@mail.ru} \\
$^1$National Nanotechnology Laboratory of Open Type, Department of Theoretical and Nuclear Physics,\\ Al-Farabi Kazakh National University, Al-Farabi ave. 71, Almaty 050040, Kazakhstan.\\
$^2$Energetic Cosmos Laboratory, Department of Physics, Nazarbayev University, Qabanbay Batyr ave. 53, Nur-Sultan 010000, Kazakhstan.
}

\begin{document}

\date{\today}
\maketitle

\begin{abstract}
Static and uniformly rotating, cold and hot white dwarfs are investigated both in Newtonian gravity and general theory of relativity, employing the well-known Chandrasekhar equation of state. The mass-radius, mass-central density, radius-central density etc relations of stable white dwarfs with $\mu=A/Z=2$ and $\mu=56/26$ (where $A$ is the average atomic weight and $Z$ is the atomic charge) are constructed for different temperatures. It is shown that near the maximum mass the mass of hot rotating white dwarfs is slightly less than for cold rotating white dwarfs, though for static white dwarfs the situation is opposite.
\end{abstract}


\section{Introduction}

White dwarfs (WDs) are the final products of the evolution of average and low-mass main sequence stars. They are formed in the cores of red giant stars. Most of the stellar population will end up as a white dwarf star. Therefore, WDs are considered to be the most abundant stellar remnants. The average mass of a white dwarf is around 0.6 solar mass ($M_\odot$) and radius is roughly 10000 km. Correspondingly, their average density is approximately 10$^6$ g/cm$^3$ \citep{1971reas.book.....Z,1971tges.book.....Z,shapirobook}.

Unlike neutron stars (NSs), there are only a few equations of state (EoS) in the literature for WDs: the Chandrasekhar EoS \citep{chandrasekhar31}, the Salpeter EoS \citep{1961ApJ...134..669S}, the Relativistic Feynman-Metropolis-Teller EoS \citep{2011PhRvD..84h4007R} and other EoS which are more sophisticated and realistic \citep{2019MNRAS.490.5839B} . These equations with some modifications, including finite temperatures, magnetic field etc. are used to describe the cores of WDs and outer crusts of NSs \citep{haenselbook}.

The physical properties of WDs have been intensively studied both in Newtonian gravity (NG) and general relativity (GR) \citep{1971tges.book.....Z,shapirobook,1968Ap......4..227S, 1971Ap......7..274A,
1990RPPh...53..837K, 2011IJMPE..20..136B, 2015mgm..conf.2468B,2018GReGr..50...38C}. It has been shown that the effects of GR are crucial to analyze the stability of WDs close to the Chandrasekhar mass limit 1.44$M_\odot$ and can be neglected for low mass WDs \citep{2011PhRvD..84h4007R}.

Here the analyzes performed in Refs.~\citep{sheyse, b2016b, boshmg18} are extended to include the effects of rigid rotation in WDs employing the Chandrasekhar EoS at finite-temperatures with $\mu=2$ and $\mu=56/26$. The mass-radius ($M-R$), mass-central density ($M-\rho$), radius-central density ($R-\rho$), mass-angular velocity ($M-\Omega$), radius-angular velocity ($R-\Omega$) and angular velocity-central density ($\Omega-R$) relations for hot static and rotating WDs at Keplerian rate are constructed both in NG and GR for comparison.

The main goal of the paper is to jointly study the effects of rotation, finite temperatures and GR only, which usually are considered separately in the literature, without involving the Thomas-Fermi corrections, Coulomb interactions between electrons and ions, phase transition etc in the EoS. Though the corrections in the EoS are important in the accurate theoretical description of the $M-R$ relations of WDs, for the sake of simplicity the Chandrasekhar EoS is used throughout this work.

\section{Problem Setup and Results}\label{sec:fopres}
The physical characteristics of rigidly rotating WDs at different finite temperatures are studied here in the range of radius from 1000 km to 200 000 km. For clarity, the Chandrasekhar EoS is exploited since it is well-known and widely used to the description of WDs \citep{chandrasekhar31,2011PhRvD..84h4007R,b2016b,2019arXiv190810806C}.

The ratio of the atomic number $A$ to the number of protons $Z$ is usually denoted in the literature as $\mu=A/Z$ and all calculations in this paper are carried out by adopting $\mu=2$ for helium $^{4}_{2}$He, carbon $^{12}_{6}$C, oxygen $^{16}_{8}$O etc and $\mu=56/26\approx2.154$ for iron $^{56}_{26}$Fe for comparison.

Helium and iron are the two extreme bounds of chemical elements in WDs. All the rest elements such as carbon, oxygen, neon, magnesium etc. are between these two, correspondingly the mass-radius relations of other elements will be restricted by helium from above and by iron from below, see Ref.~\citep{2011PhRvD..84h4007R} for details.

Furthermore, WDs composed of helium were observed and their formation were analyzed in simulations with different astrophysical scenarious \citep{2001AN....322..405S, 2004ApJ...606L.147L, 2005MNRAS.362..891B, 2014A&A...571L...3I}. There are also a plethora of observational data for carbon and other WDs \citep{kepler2015, kepler2016}. The time and thermal evolution of carbon WDs has been theoretically investigated involving the nuclear burning and neutrino emission processes \citep{2019MNRAS.487..812B}. The observational support of the existence of iron-rich cores of WDs was given in Ref. \citep{1998ApJ...494..759P}. The structure and evolution of iron-core WDs has been studied in Ref. \citep{2000MNRAS.312..531P}. Some simulations demonstrate that the formation mechanism of WDs with iron-rich cores are different from the WDs composed of light elements \citep{2012ApJ...761L..23J}.

For the construction of $M-R$ relations the temperatures of the WD isothermal cores $T=T_c$ are considered here without taking into account the atmosphere of WDs. The  effective surface temperature $T_{eff}$, what is usually measured from observations, is roughly three order of magnitude less than $T_c$ according to the approximate Koester formula  $T_{eff}^4/g=2.05\times10^{-10} T_c^{2.56}$, where $g$ is the surface gravity \citep{koester2}.

To investigate static WDs in GR the Tolman-Oppenheimer-Volkoff (TOV) hydrostatic equilibrium equations are solved numerically. To construct rotating WDs the Hartle formalism is applied both in GR and NG \citep{Hartle,1968ApJ...153..807H, bosh2016} which generalizes the TOV equations by the inclusion of rotation. Note, that Ref~\citep{bosh2016} is a pedagogical paper explicitly showing the Hartle approach formulated in NG and comparing with other treatments in the literature in detail. All rotating WDs in NG are calculated at the Keplerian sequence: $\Omega_{Kep}=\sqrt{G M_{rot}/R^3_{eq}}$, where $\Omega_{Kep}=\Omega$ is the maximum angular velocity, $M_{rot}$ is the total rotating mass, and $R_{eq}$ is the corresponding equatorial radius of WDs. In GR the corrections to $\Omega_{Kep}$ owing to the angular momentum (the Lense-Thirring effect) and quadrupole moment (deformation of a star) are duly taken into account.

It should be stressed that though the Hartle formalism is designed and valid for slowly rotating stars in the approximation of $\sim \Omega^2$ (angular velocity of a star) one can extrapolate it to the Keplerian mass shedding limit for qualitative analyses. Indeed, Hartle and Thorne \citep{1968ApJ...153..807H} applied the approach for rapidly rotating stars (from supermassive stars to NSs) to study the effects of rotation on the structure. The validity and reliability of the formalism was tested in Refs.~\citep{berti2004,berti2005}. It was shown that the discrepancies between the Hartle formalism and exact computations appear close to the mass shedding limit. In addition, it has been demonstrated that the so-called Sedrakyan and Chubaryan formalism \citep{1968Ap......4..227S, 1971Ap......7..274A}, usually applied to the  study of rigidly rotating WDs and NSs, is identical with the Hartle formalism \citep{2016GrCo...22..305B}. Therefore, one can safely employ the Hartle formalism at the  mass shedding limit for rough estimates and qualitative analyses \citep{2000csnp.conf.....G,stergioulas}.

One can also study the effects of differential rotation, contribution of the atmosphere, diffent temperature profiles ect inside WDs in analogy with Refs. \citep{2019MNRAS.486.2982Y,2020MNRAS.492..978T}. However, those issues are out of the scope of the current work.

\begin{figure*}
\centering
{\includegraphics[width=3.2in]{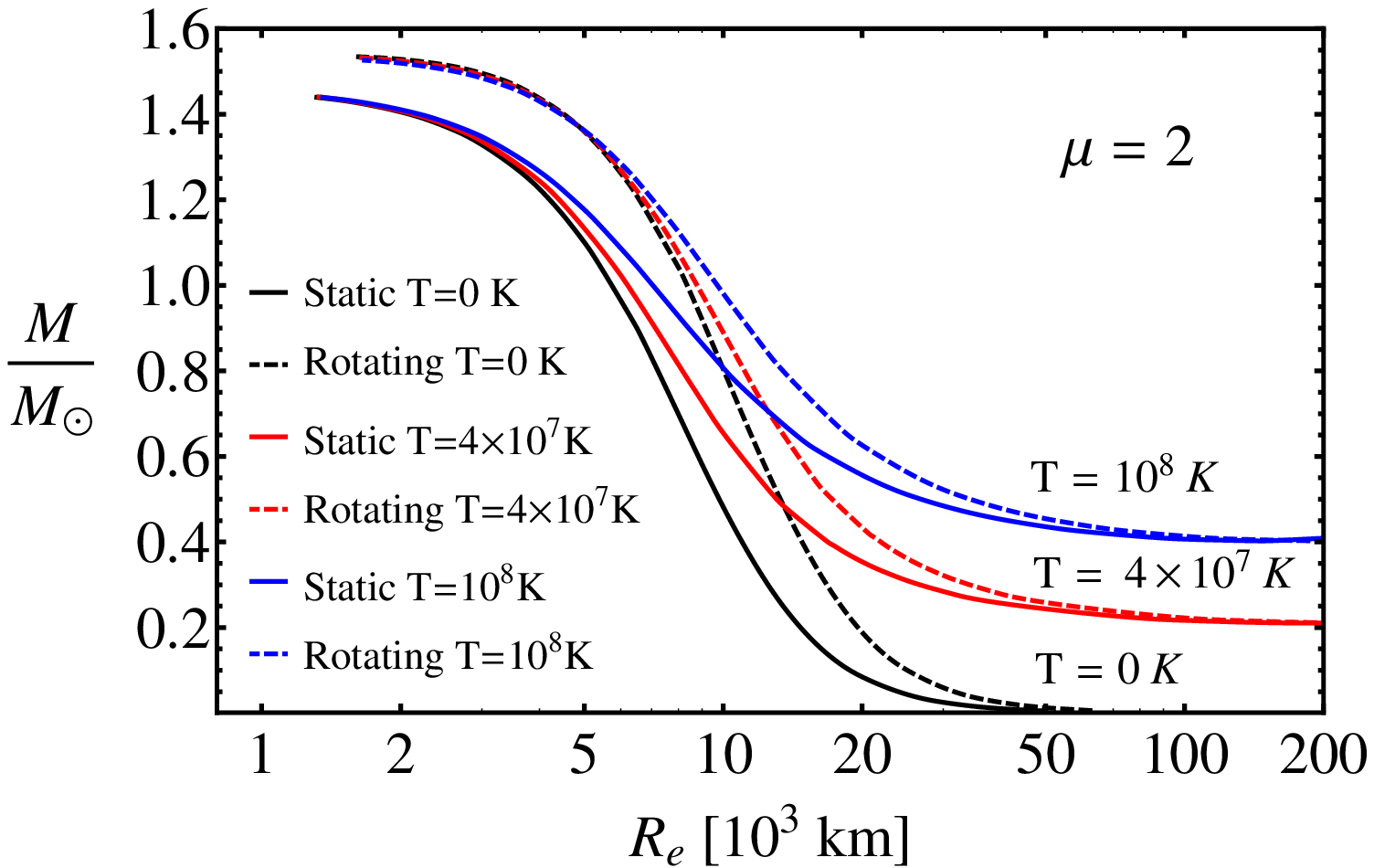}}
{\includegraphics[width=3.2in]{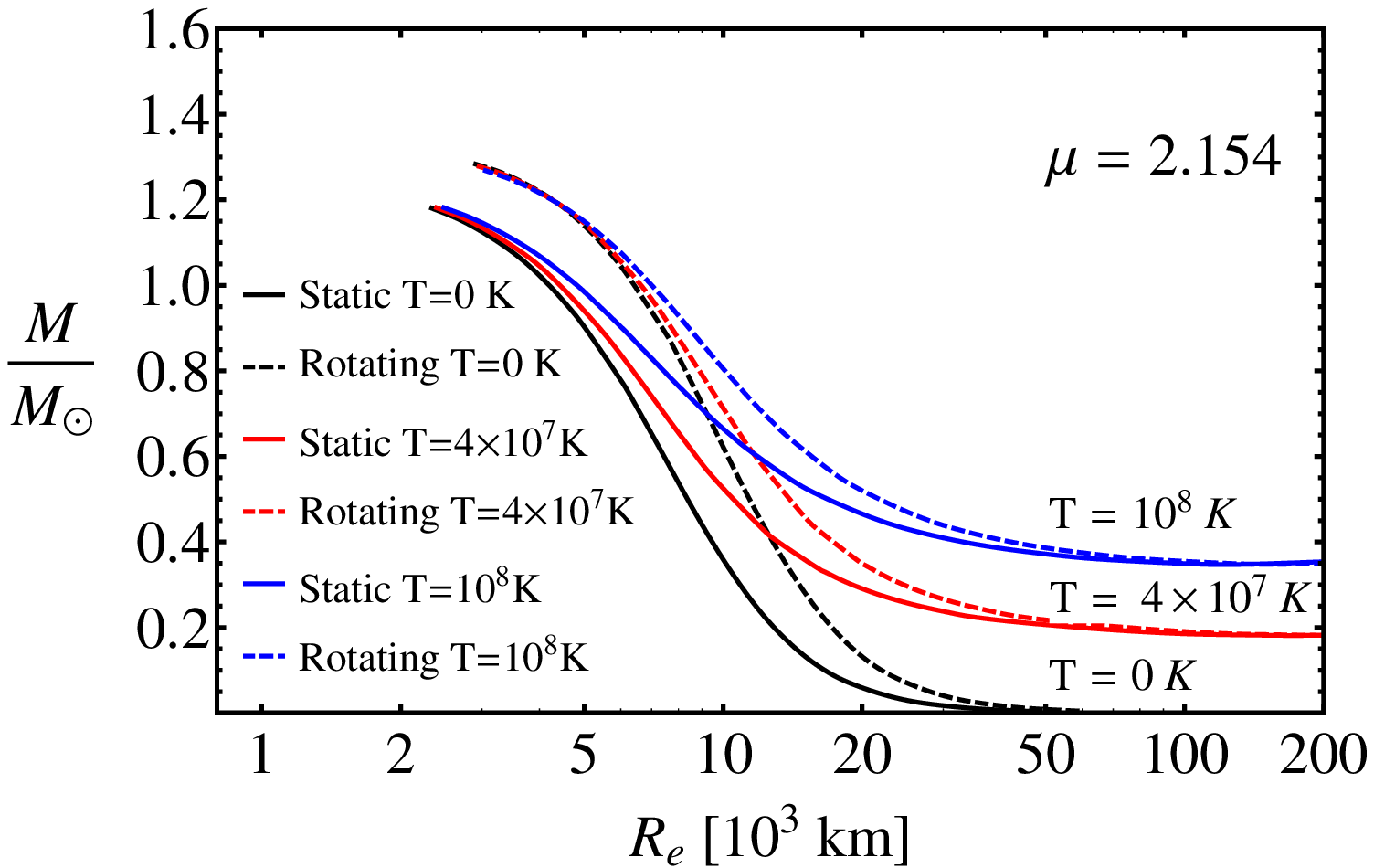}}
\caption{Colour online. Mass-radius relations for WDs with $\mu=2$ (left panel) and for WDs with $\mu=2.154$ (right panel). In both figures the solid curves show static and the dashed curves show rotating WDs for different temperatures T=[0, 4$\times10^7$, $10^8$] K.}\label{fig:f1}
\end{figure*}

By solving the structure equations numerically,  $M-R$, $M-\rho$, $R-\rho$, $M-\Omega$, $R-\Omega$ and $\Omega-\rho$ relations for static and rotating hot WDs are obtained. The stability of hot rotating WDs has been analyzed in Refs.~\citep{bosh2018,bosh2018AR}. In Fig.~\ref{fig:f1} the $M-R$ relations are shown at different temperatures $T=[\ 0, 4\times10^7, 10^8]\ $K depicted with black, red and blue curves, respectively, for $\mu=2$ (left panel) and $\mu=2.154$ (right panel). All solid curves indicate static WDs and dashed curves correspond to rotating WDs. It is apparent that the finite-temperature effects are more prominent for larger in size (low mass) WDs and the effects of rotation are crucial for smaller in size (massive) WDs. Nevertheless both the finite temperature and rotation effects contribute to the radius and the mass of WDs. The comparison of these theoretical curves for $\mu=2$ with the estimated masses and radii of WDs from the Sloan Digital Sky Survey Data Release 4 (see Ref.~\citep{tremblay}) is given in Ref.~\citep{bosh2018AR}.

The behavior of the $M-R$ relations with $\mu=2$ and $\mu=56/26$ is similar, though the mass is larger for WDs with $\mu=2$. One can see that for a fixed mass within the range of the equatorial radius $R_{eq}=(5\times10^3-50\times10^3)$ km of Fig.~\ref{fig:f1} the higher the temperature the larger the radii. The same is true for a fixed radius: the higher the temperature the larger the mass.

\begin{figure}
\centering
\includegraphics[width=1\columnwidth,clip]{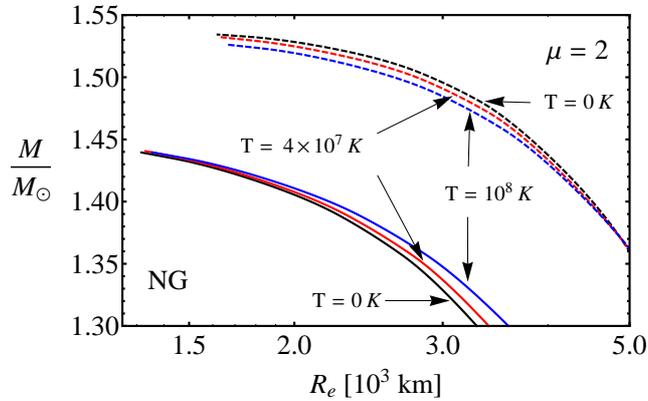}
\caption{Colour online. Mass-radius relations of WDs close to the maximum mass.}\label{fig:f2}
\end{figure}

Fig.~\ref{fig:f2} is magnified Fig.~\ref{fig:f1} (left panel) in the range $R_{eq}=(1.1\times10^3-5\times10^3)$ km. Previously the range near the maximum mass was not analyzed thoroughly, albeit the general behavior of the $M-R$ relations was known \citep{bosh2018,bosh2018AR}. Within that range for static WDs, the higher temperatures the larger masses. This effect is natural, since due to the temperature, for a fixed central density, the pressure of partially degenerate and non-degenerate electrons increases and can sustain more mass. Instead, for rotating WDs, the higher temperatures the smaller masses. This effect is counter-intuitive, since one would expect that the effects observed in static WDs would automatically translate to rotating WDs. However, this is not the case, at least in this range of mass and radius. Here, the temperature affects the radius more than the mass. Hence, the Keplerian angular velocity is lower for hotter WDs, correspondingly the total rotating mass is less than for colder WDs.

\begin{figure*}
\centering
{\includegraphics[width=3.2in]{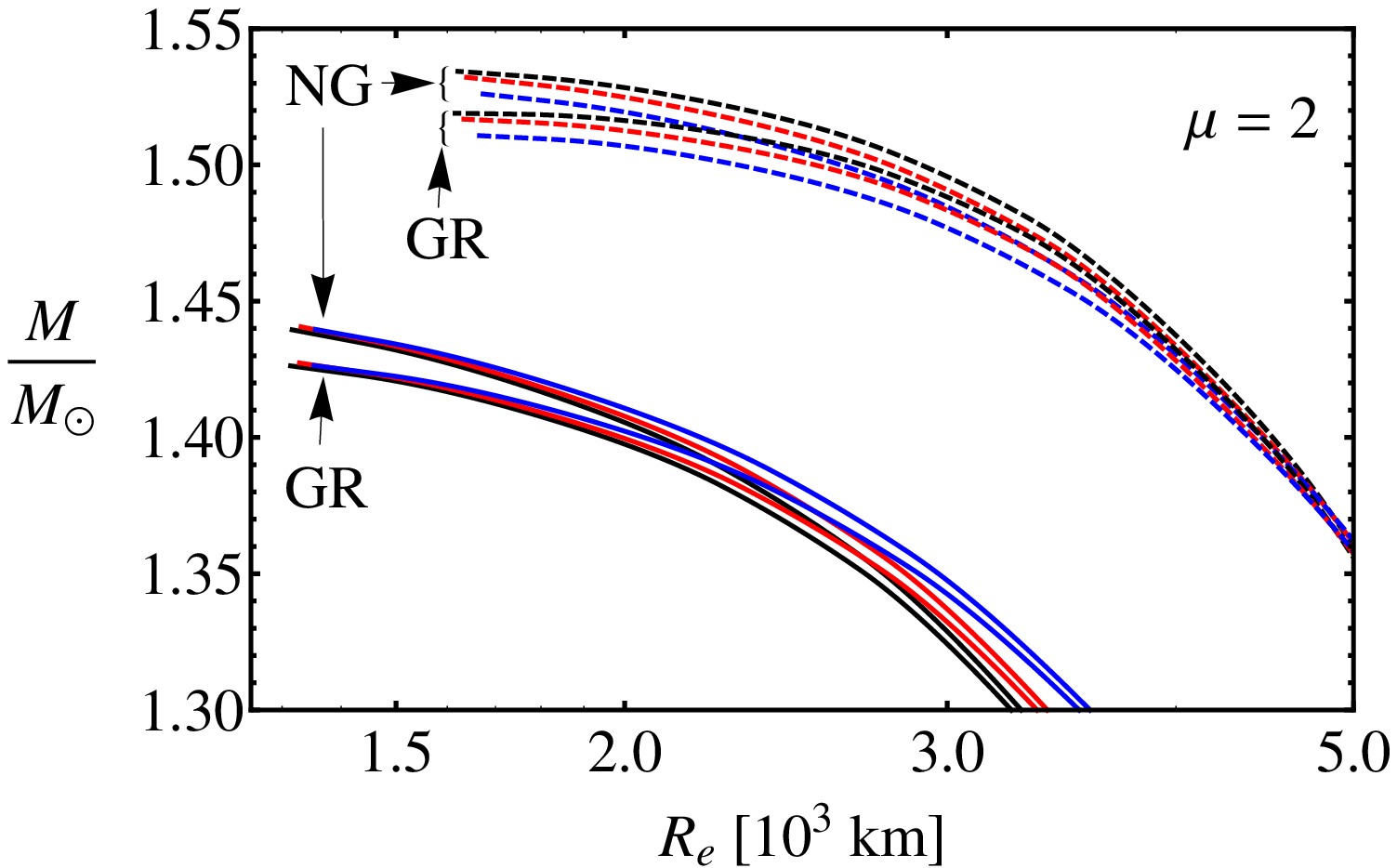}}
{\includegraphics[width=3.2in]{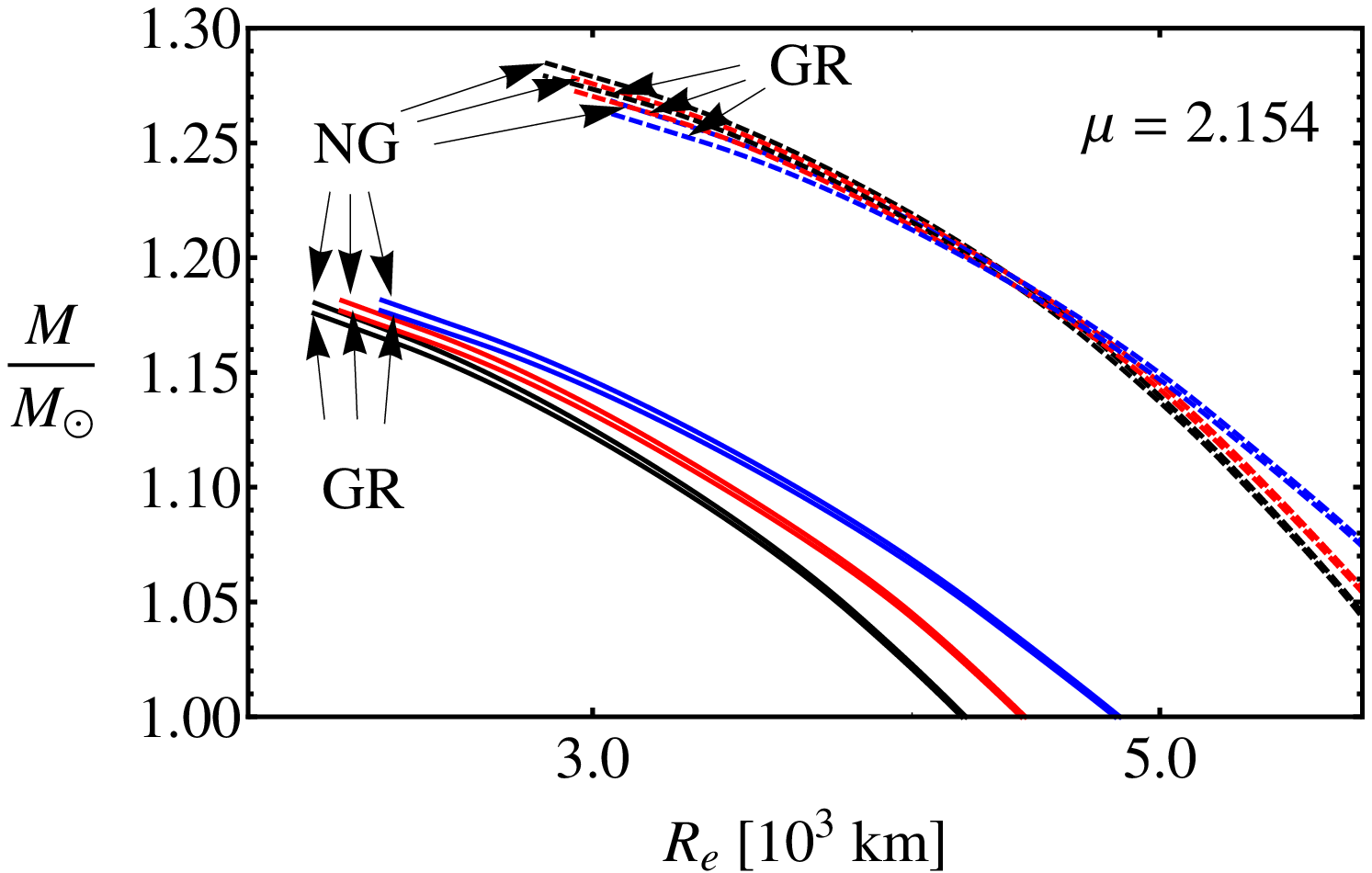}}
\caption{Colour online. Mass-radius relations of WDs close to the maximum mass.}\label{fig:f222}
\end{figure*}

Fig.~\ref{fig:f222} shows $M-R$ relations close to the maximum mass both in NG and GR. As one can see that for the $\mu=2$ case (left panel) the role of GR is crucial. Instead GR is negligible for the $\mu=56/26$ case (right panel).

\begin{figure*}
\centering
{\includegraphics[width=3.2in]{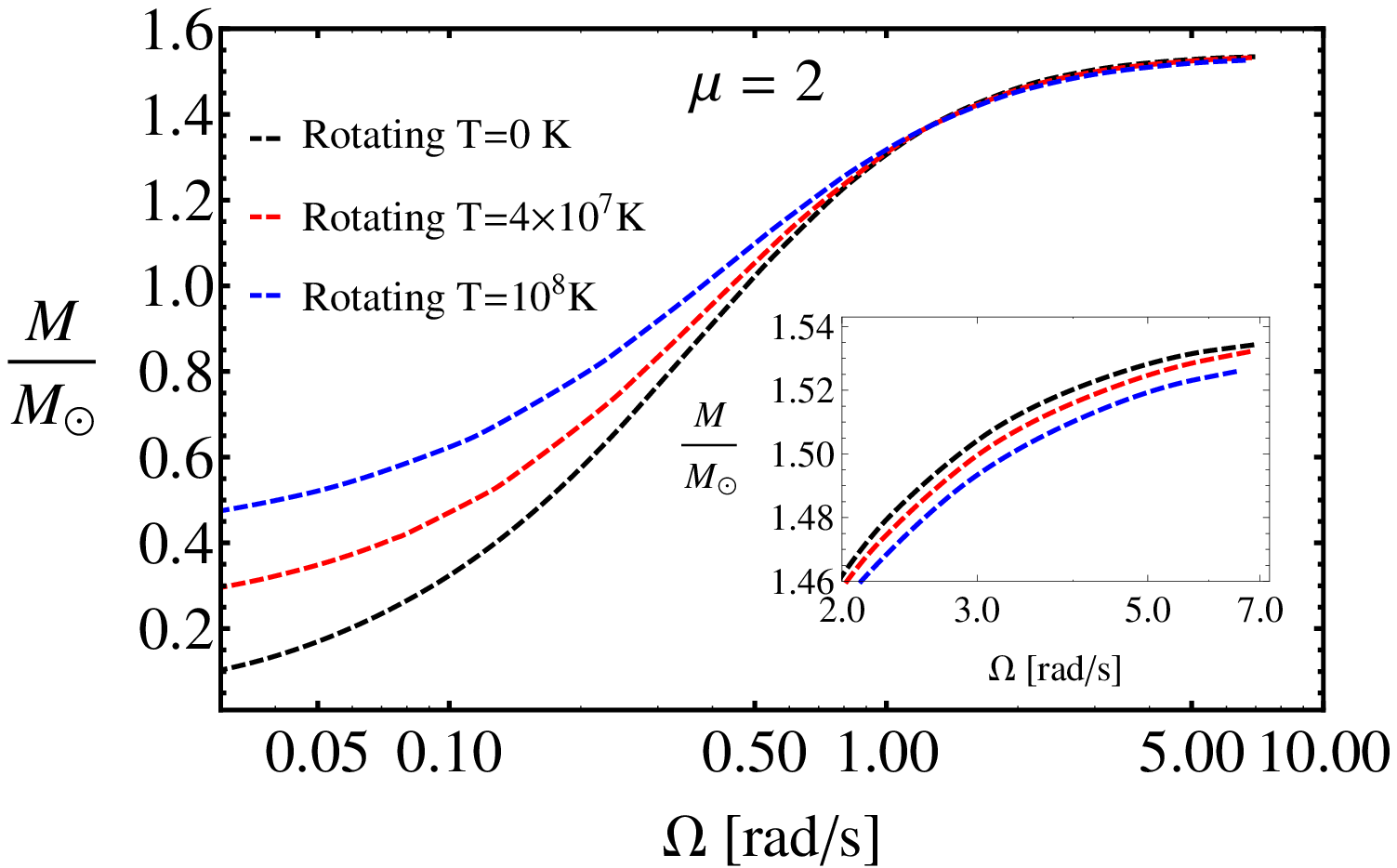}}
{\includegraphics[width=3.2in]{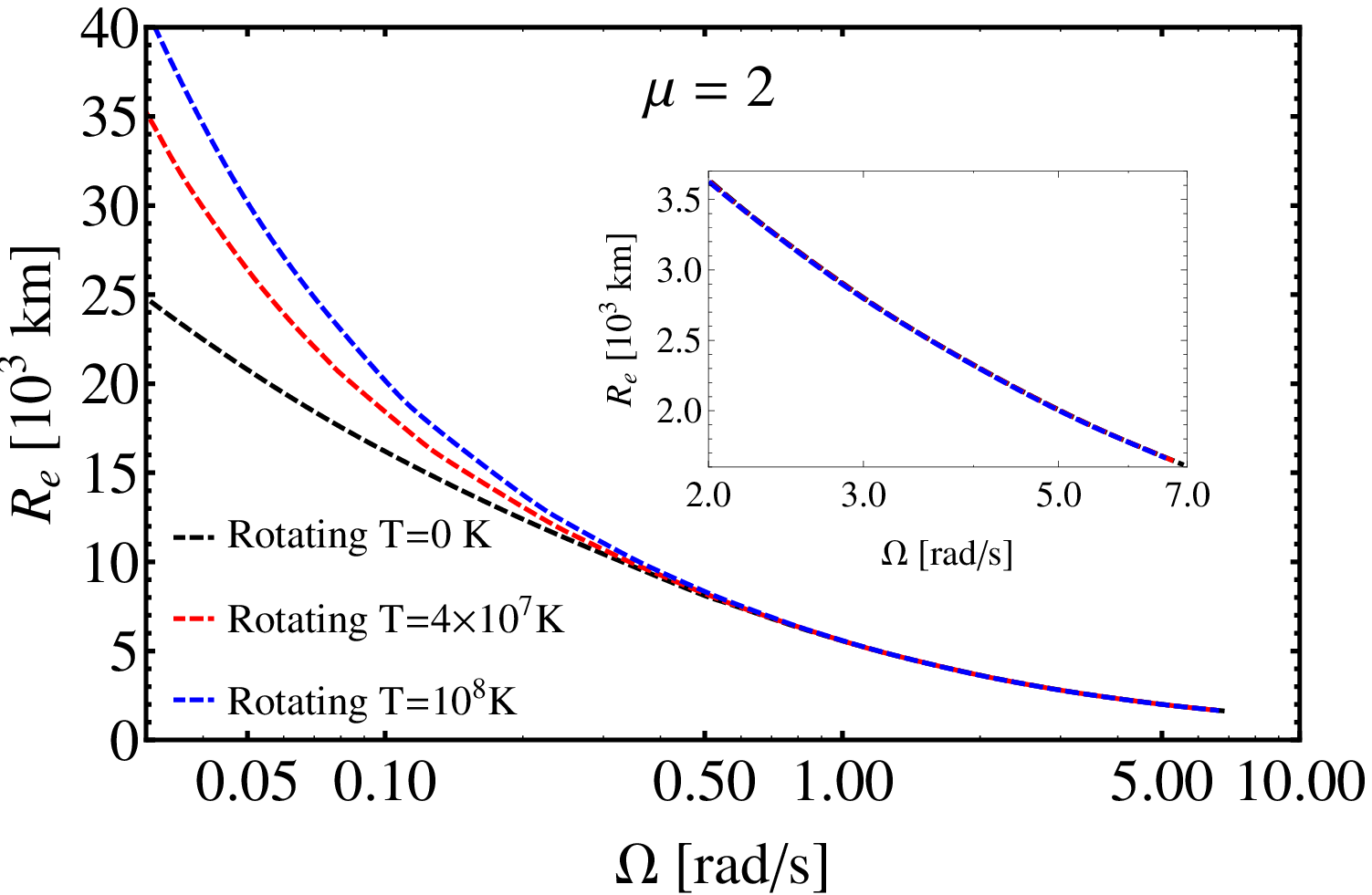}}
\caption{Colour online. Mass-angular velocity relations (left panel) and equatorial radius-angular velocity relations (right panel) of WDs close to the maximum mass.}\label{fig:f3}
\end{figure*}

In Fig.~\ref{fig:f3} (left panel) the dependence of the total mass is constructed as a function of the angular velocity. It is clear that for a fixed mass cold WDs rotate faster. However the situation is opposite close to the maximum mass. Right panel of Fig.~\ref{fig:f3} shows the dependence of the equatorial radius on the angular velocity. Here for a fixed radius hot WDs rotate faster and close to the maximum mass the rotation rate of WDs is almost independent of the temperature.

\begin{figure*}
\centering
{\includegraphics[width=3.2in]{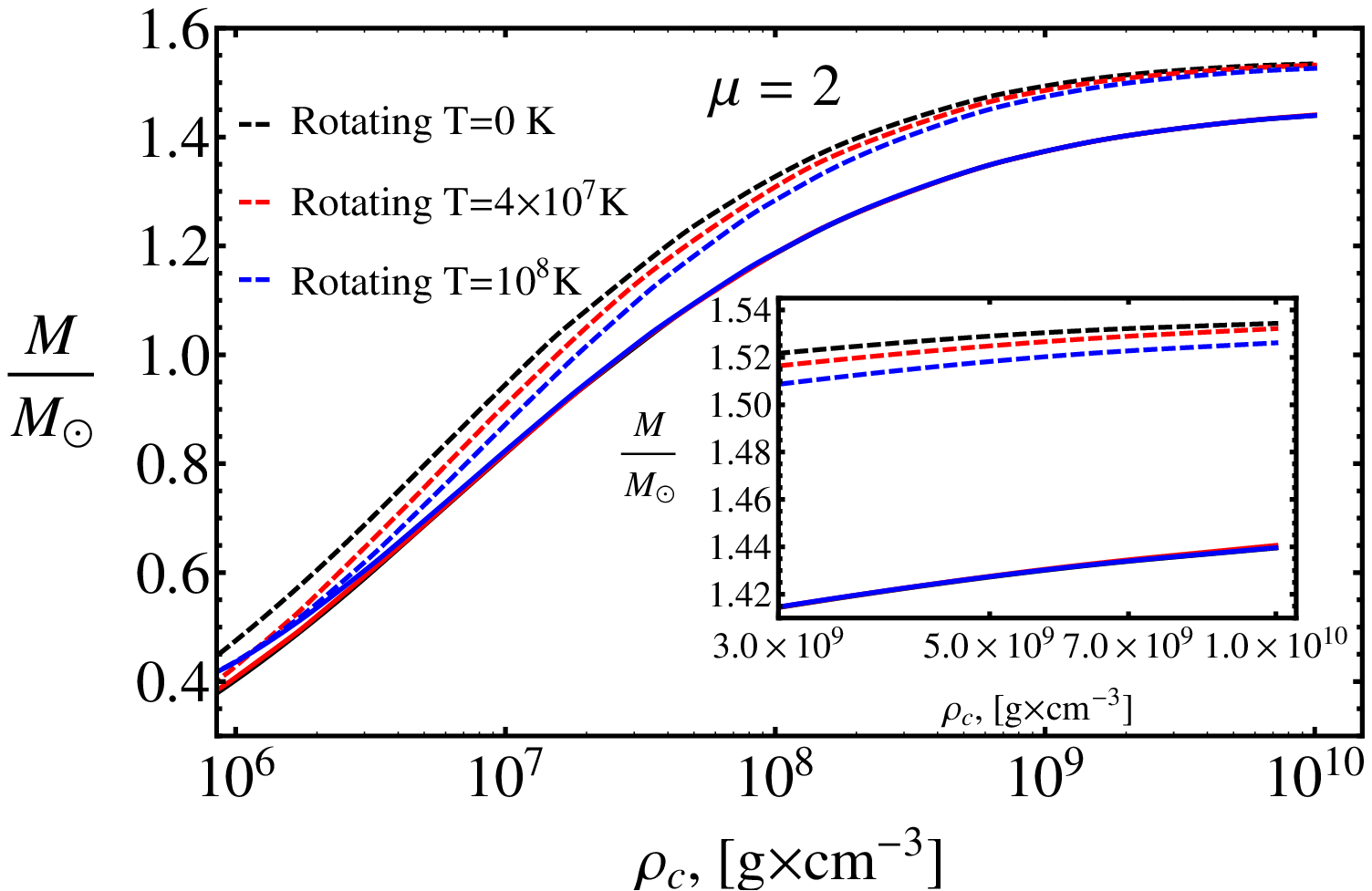}}
{\includegraphics[width=3.2in]{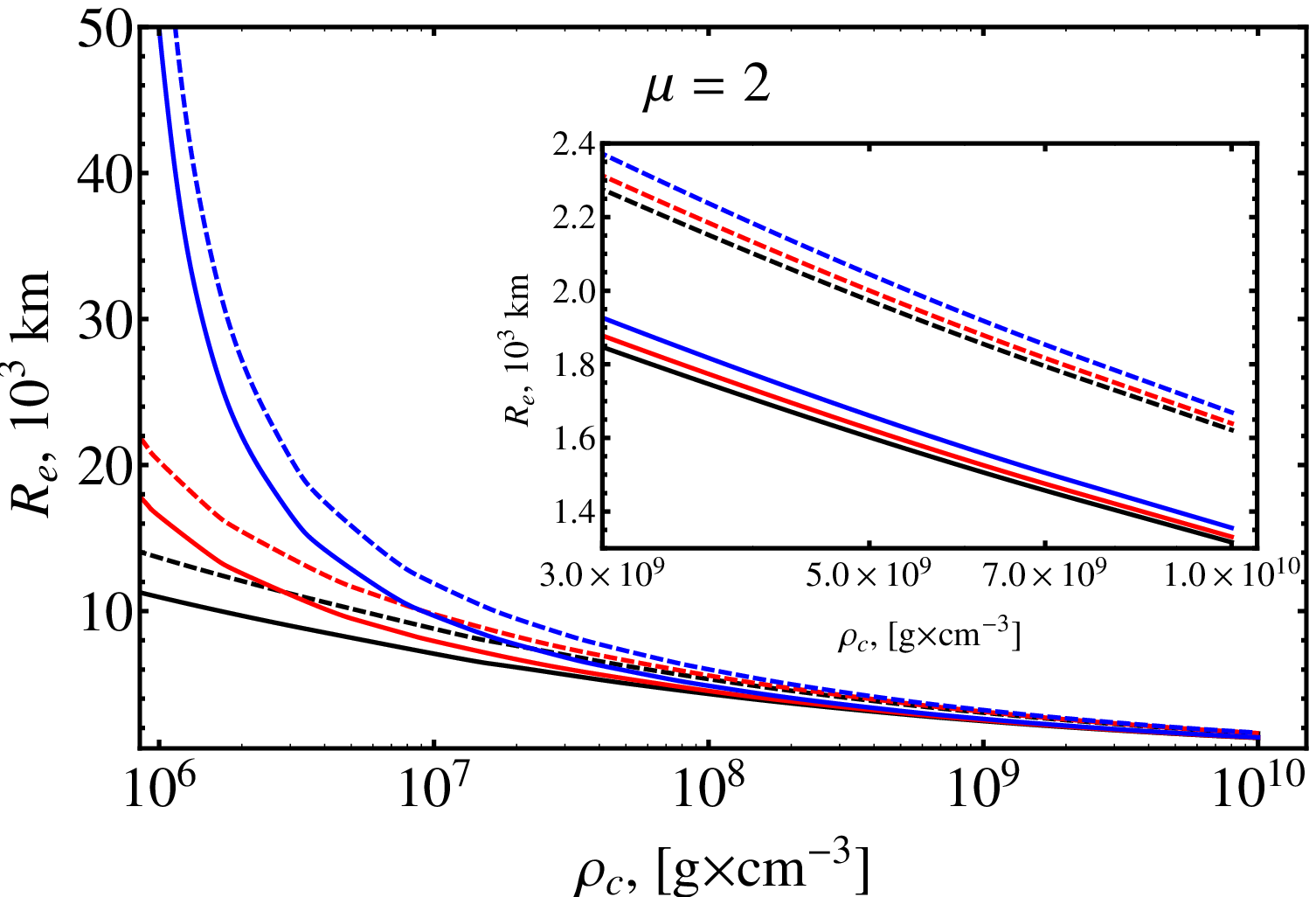}}
\caption{Colour online. Mass-central density relations (left panel) and equatorial radius-central density relations (right panel) of rotating cold and hot WDs.}\label{fig:f4}
\end{figure*}

To better understand the combined effects of rotation and finite temperatures on WDs $M-\rho$ and $R-\rho$ relations are constructed in Fig.~\ref{fig:f4}. Left panel of Fig.~\ref{fig:f4} shows that for a fixed central density rotating hot WDs in the considered density range will be less massive than colder ones. On the contrary for a fixed central density the radius of hot WDs will be always larger than the colder ones (see Fig.~\ref{fig:f4} (right panel)).

\begin{figure}
\centering
\includegraphics[width=\columnwidth,clip]{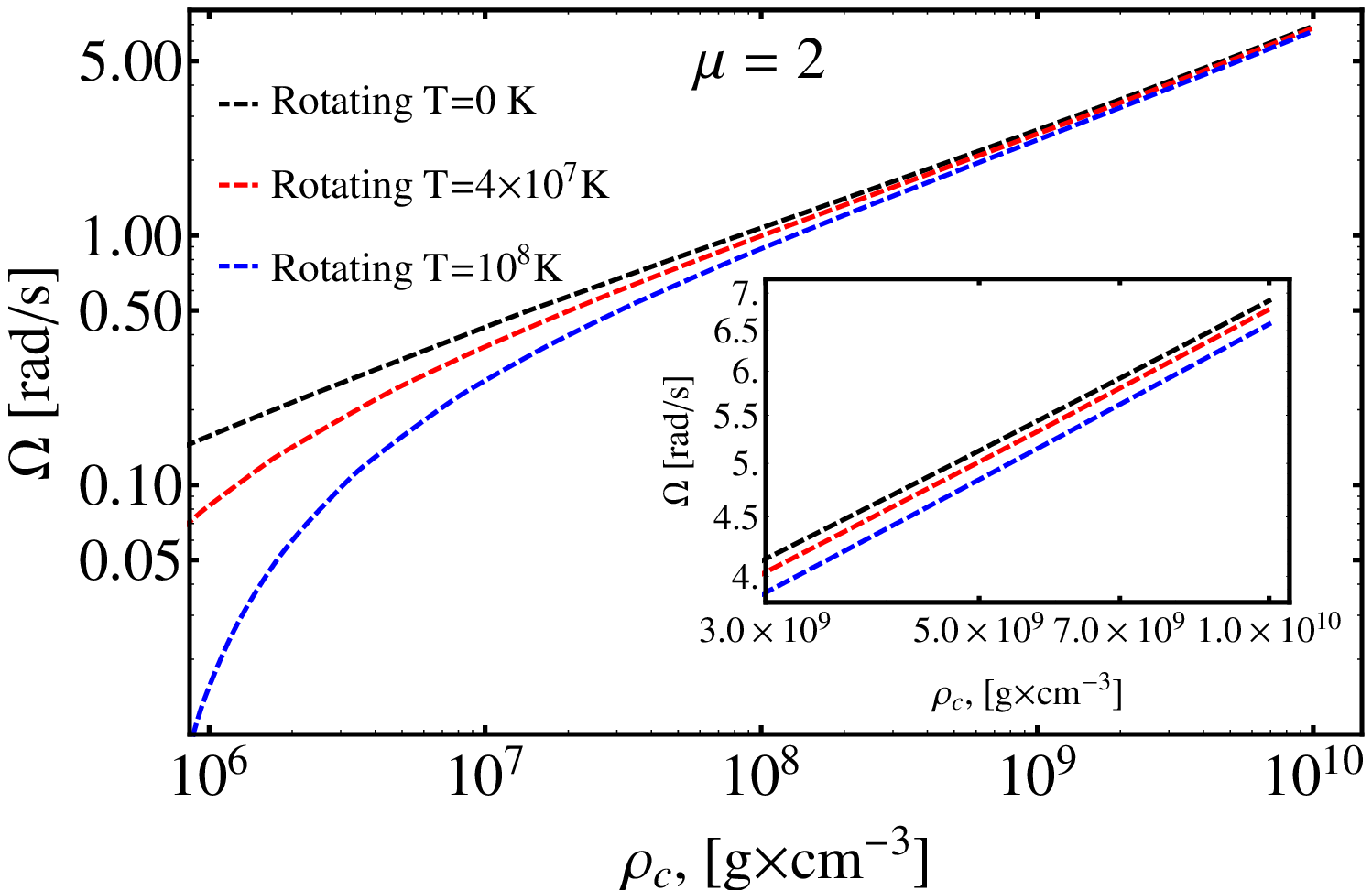}
\caption{Colour online. Angular velocity versus central density for rotating WDs.}\label{fig:f5}
\end{figure}

According to Fig.~\ref{fig:f5} for a fixed central density colder WDs will always rotate faster than hotter ones.

\begin{table*}
\centering
 \caption{Static and rotating masses for hot white dwarfs for fixed central densities both in Newtonian gravity and in general relativity. The values of the mass in general relativity are given in parentheses.}
 \label{tab:example}
 \begin{tabular}{lccccccc}
  \hline
  & Configurations &$T=0$ K & $T=4\times10^7$ K & $T=10^8$ K \\
  \hline
$M/M_{\odot}$,  $\mu=2$ & Static & 1.440 (1.426) & 1.440 (1.426) & 1.440 (1.426) \\

$\rho_c=10^{10}$ g/cm$^3$& Rotating & 1.534 (1.519) & 1.532 (1.517) & 1.526 (1.511)\\
 \hline
 $M/M_{\odot}$,  $\mu=56/26$ & Static  & 1.180 (1.176) & 1.181 (1.177) & 1.181 (1.177) \\

$\rho_c=10^9$ g/cm$^3$ & Rotating  & 1.284 (1.279) & 1.278 (1.273) & 1.268 (1.263)\\
  \hline
 \end{tabular}
\end{table*}

In Table~\ref{tab:example} the static and rotating masses of WDs are presented corresponding to $\mu=2$ with the central density $\rho_c=10^{10}$ g/cm$^3$ and $\mu=56/26$ with $\rho_c=10^9$ g/cm$^3$. The values of the critical central density are chosen in accordance with the inverse $\beta$-decay and pycnonuclear reaction (for carbon WDs) instability densities, see Ref.~\citep{2013ApJ...762..117B} for details. As one can see that for static WDs the maximum mass almost does not change with the increasing temperature (see Fig.~\ref{fig:f4}), instead for rotating white dwarfs the maximum mass slightly decreases with the increasing temperature in both $\mu=2$ and $\mu=56/26$ cases.

\begin{table*}
\centering
 \caption{Static and rotating masses for hot white dwarfs for a fixed radius both in Newtonian gravity and in general relativity. The values of the mass in general relativity are given in parentheses.}
 \label{tab:example2}
 \begin{tabular}{lccccccc}
  \hline
  & Configurations &$T=0$ K & $T=4\times10^7$ K & $T=10^8$ K \\
  \hline
$M/M_{\odot}$,  $\mu=2$ & Static & 1.329 (1.324) & 1.337 (1.332)& 1.347 (1.342)\\

$R_{eq}=3\times10^{3}$ km & Rotating & 1.495 (1.488) & 1.490 (1.483) & 1.484 (1.477)\\

  \hline
 \end{tabular}
\end{table*}

In Table~\ref{tab:example2} total static and rotating masses are given for a fixed equatorial radius (3000 km) at different temperatures in $\mu=2$ case only. Note that for a static case the equatorial radius reduces to the static radius. It is again evident that rotating hot WDs will possess less mass with respect to cold ones, in accordance with Fig.~\ref{fig:f2}. In $\mu=56/26$ case this effect is negligible due to the limitation in the central density caused by the inverse $\beta$-decay instability.

\section{Discussion and Conclusion}\label{sec:con}
The properties of static and rotating WDs have been investigated using the Chandrasekhar EoS at finite-temperatures. The structure equations have been solved numerically to construct $M-R$, $M-\rho$, $R-\rho$, $M-\Omega$, $R-\Omega$ and $\Omega-\rho$ relations for hot static and rotating isothermal cores of WDs. The atmosphere of WDs was not considered for simplicity.

It has been shown that temperature affects the masses of WDs at larger radii. At smaller radii the thermal effects are negligible. Rotation affects the masses of cold WDs in all density range. For hot WDs the effects of rotation are less noticeable at larger radii and more noticeable at smaller radii.

The $M-R$ relations were compared and contrasted for cold and hot, static and rotating WDs with $\mu=2$ and $\mu=56/26$. If the case with $\mu=2$ was well studied in the literature, the case with $\mu=56/26$ including the effects of both rotation and finite temperatures is investigated here for the first time. It turned out that for a fixed mass hot iron WDs are smaller in size, correspondingly denser with respect to the WDs composed of light elements.

It was found that close to the maximum mass the hotter rotating WDs possess less mass than the colder ones. This is related to the fact that unlike in the static case where the  mass of a WD is a function of the central density, radius and temperature, in the rotating case one more variable i.e. angular velocity is involved in a non-trivial way. Therefore, one observes that within the range $R_{eq}=(1\times10^3-5\times10^3)$ km  hot rotating WDs posses slightly less mass than the cold ones. To the best knowledge of the author this result is new.

In addition, to better comprehend the decrease in mass of hot rotating WDs it was useful to construct $M-\rho$, $R-\rho$, $M-\Omega$, $R-\Omega$ and $\Omega-\rho$  relations. Since in our case angular velocity is an additional degree of freedon its effects to the structure of WDs along with temperature are not straightforward.

In summary, one can see in Fig.~\ref{fig:f4} that for hot {\it static} WDs the maximum mass does not change (left panel) with increasing temperature, this is related to the Fermi temperature which is very high at this density range, instead the corresponding {\it static} radius (right panel)   changes: the hotter the WD the larger its radius. Hence, hotter WDs are fluffier than the colder ones. Consequently, larger in size hot WDs cannot rotate faster than smaller in size cold WDs with the same central density. If they cannot rotate faster, in order to fulfill the stability criteria i.e. the central density must not exceed a certain value; their rotating total mass must be lower than for the colder WDs.

Furthermore, for the sake of comparison all the computations have been performed both in NG and GR. As it was expected the effects appeared in NG are automatically translated into GR. The contribution of GR to the $M-R$ becomes relevant close to the maximum mass for $\mu=2$ cases. As for the $\mu=56/26$ case the effects of GR are negligible though are crucial for stability analysis and the correct estimation of the radius. Therefore the main focus was given to the $\mu=2$ case.

In view of the latest observational data on WDs \citep{kepler2019, 2019MNRAS.482.4570G} it would be interesting to explore WDs, taking into account the Coulomb interactions and Thomas-Fermi corrections at finite temperatures in the EoS as in Ref.~\citep{sheyse,Faussurier2017,2017JSMTE..11.3101F}, including the effects of rotation in the structure equations. That will be the issue of future studies.


\section*{Acknowledgement}
\medskip
\noindent
This work was supported by the Ministry of Education and Science of the Republic of Kazakhstan, Program IRN: BR05236494, Grants IRN: AP05135753  and IRN: AP08052311 . 

\bibliographystyle{mnras}

\end{document}